# Realization of ideal flat band by rotated d-orbitals in Kagome metals

Dongwook Kim and Feng Liu[1]

*Department of Materials Science and Engineering, University of Utah, Salt Lake City, Utah 84112, USA*

Abstract

Recently there has been intense interest in Kagome metals, which are expected to host flat bands (FBs). However, the observed FBs are non-ideal as they are not flat over the whole 2D Brillouin zone and overlap strongly with other bands. Most critically, the theoretical conditions for the existence of ideal FB in Kagome metals, beyond the simplest Kagome lattice model, are unknown. Here, based on tight-binding model analyses of the interplay between *d*-orbital symmetry and underlying Kagome lattice symmetry, we establish such conditions. We show that for a pure transition-metal (TM) Kagome lattice, only $d_{z^2}$ orbital gives rise to a FB; while $d_{xy}$, $d_{x^2-y^2}$, $d_{xz}$, and $d_{yz}$ orbitals do not. The condition for the latter four orbitals to produce a FB is to rotate them so that they will conform with the underlying Kagome lattice symmetry. Interestingly, the lattice having rotated $d_{xy}$ ($d_{xz}$) and $d_{x^2-y^2}$ ($d_{yz}$) orbitals leads to a FB of opposite chirality sitting above and below the Dirac bands, respectively. For intercalated TM Kagome lattices, the Kagome-hexagonal intercalation exhibits always a FB, while the case for the Kagome-trigonal intercalation is conditional. Another condition to isolate the FB arising from individual *d*-orbitals is to increase the crystal field splitting, such as in Kagome metals of layered metalorganic frameworks.

---

[1] fliu@eng.utah.edu

*Introduction.*—Kagome lattice is arguably the most intriguing lattice. A spin Kagome lattice is a prominent candidate for quantum spin liquid (QSL) due to geometrical frustration, while the quasiparticle eigen spectra of a Kagome lattice contains an eigenvalue with macroscopic degeneracy, a flat band (FB), due to destructive quantum interference, i.e., phase cancellation of Bloch wave function. Hybrid Kagome metals, namely compounds containing layers of Kagome sublattice of transition metals (TM) sandwiched by layers of organic ligands have been long investigated to search for signatures of QSL [1–6], as well as other magnetic quantum states, such as quantum optical spin ice [7–11], Kagome magnet [4,12,13], anomalous Hall effect [14] and skyrmion [15].

Recently, inorganic Kagome metals, such as $CoSn$, $Fe_3Sn_2$, $CsV_3Sb_5$, $YCr_6Ge_6$, and $Ni_3In$, have drawn increasing attention, due to the presence of FBs and Dirac bands with von Hove singularities, which lead to a range of observed interesting physical phenomena, such as ferromagnetism [16–20], superconductivity [21–24]. However, the experimentally observed FBs [25–29] as well as the density-functional-theory (DFT)-calculated band structures are non-ideal, as the presumed "FBs" are not flat over the whole 2D Brillouin zone and buried with strong overlap with many other bands around the Fermi level [25–29]. Moreover, there remain fundamental gaps in our understanding of *d*-orbital FBs in Kagome metals. Most critically, the physical underpinning for the existence of ideal FB in Kagome metals, beyond the simplest Kagome lattice model assuming an *s*-orbital per lattice site, is largely unknown. Therefore, it is highly desirable to establish a *d*-orbital model including anisotropic orbital symmetries to better understand the emergence of FB in Kagome metals, which will open a promising avenue to realizing some elusive predicted FB phenomena, such as the fractional QHE [30–35], Wigner crystallization [36–39], excitonic insulator [40,41], QAH/ QSH effect [42–45].

In this Letter, we develop a full tight-binding (TB) *d*-orbital Kagome lattice model, to establish the general physical principles for the existence of ideal FB in Kagome metals, by revealing the important interplay between *d*-orbital symmetry and underlying Kagome lattice symmetry. We show that for a Kagome metal consisting of pure 2D TM Kagome lattice planes, only $d_{z^2}$ orbital gives rise to a FB; while $d_{xy}$, $d_{x^2-y^2}$, $d_{xz}$, and $d_{yz}$ orbitals do not. The condition for the latter four orbitals to produce a FB is to rotate them so that they become conform the Kagome lattice symmetry, most notably C$_3$ rotation plus translation (C$_3$+T). Interestingly, the lattice having rotated $d_{xy}$ ($d_{xz}$) and $d_{x^2-y^2}$($d_{yz}$) orbitals leads to a FB of opposite chirality sitting above and below the Dirac bands, respectively. For intercalated TM Kagome lattice planes, the Kagome-hexagonal intercalation exhibits always a FB, while the case for the Kagome-trigonal intercalation is conditional depending on the interaction between the two sublattices. Another condition to isolate the FB arising from individual *d*-orbitals is to increase the crystal field splitting (CFS). It turns out that all the currently known inorganic Kagome metals satisfy the condition of *d*-orbital rotation but having too weak a CFS. Furthermore, we propose layered metalorganic frameworks (MOFs) to be a family of Kagome metals hosting ideal FB.

*Tight-binding model of rotated d-orbitals in Kagome lattice—* It is important to recognize that the basic Kagome lattice model assumes by default a single *s* or $p_z$ orbital of even parity sitting at each lattice site [16,46–50]. The FB arises from purely lattice symmetry, such as underlined by line-graph theorem [51–54]. When *d*-orbitals are placed on each lattice site, however, several complications have to be considered. First, the five *d*–orbitals have distinct symmetries, which do not necessarily conform the underlying Kagome lattice symmetry. Secondly, unlike the *s-s* orbital hopping that is isotropic, inter-*d*-orbital hopping is directional dependent, which inevitably affects the existence of FB. Thirdly, the atomic TM *d*-orbitals have five-fold degeneracy, which will be lifted by CFS, splitting their on-site energies. When the CFS is weak, the inter-*d*-orbital hopping changes each individual subset of *d*-bands and causes overlap between them. Therefore, the existence of FB in Kagome metals is rather nontrivial, much beyond the commonly perceived simple Kagome lattice model.

To concretely illustrate the above points, we develop a full TB *d*-orbital Kagome lattice model, by explicitly implementing the five *d*-orbital symmetries to calculate band structure. Figure *1*(a) shows the schematic diagrams of a Kagome lattices having the $d_{z^2}$, $d_{x^2-y^2}$, $d_{xy}$, $d_{zx}$, and $d_{yz}$ orbital, as examples, placed at each lattice site in the default *d*-orbital orientation, respectively. Figure *1*(b) shows the corresponding calculated band structures. One sees that only the $d_{z^2}$-orbital Kagome lattice produces a perfect FB, while all other four orbitals fail. This is because the inter-$d_{z^2}$-orbital hopping within the 2D plane is isotropic. In other words, the $d_{z^2}$ orbital symmetry conforms the underlying Kagome lattice symmetry, same as for *s*- or $p_z$-orbital. In contrast, the other four orbitals have a two-fold rotation symmetry which does not conform the underlying lattice symmetries, e.g., $C_3$+T, and the interatomic hopping between them is anisotropic and directional dependent. Consequently, these four *d*-orbital symmetries interfere with the Kagome lattice symmetry to disrupt the condition of phase cancellation of Bloch wavefunction [55] and hence to mitigate the FB.

We note that in a Kagome lattice with default *d*-orbital orientation without rotation [Fig. 1 (a)]. We used typical hopping strength ($V_{dd\sigma} = -1.20t_0$, $V_{dd\pi} = +0.90t_0$, and $V_{dd\delta} = -0.10t_0$) for TM metals [56] in the Slater-Koster formalism [57] to calculate the bands [Fig. 1(b)]. Due to the nature of localized *d*-orbitals, bandwidths are generally narrow and some appears rather "flat" [see middle band in the last two columns of Fig. 1(b)]. But they are isolated bands, different from the *topological* FB hosted in the Kagome lattice, which has a singular band touching point with a dispersive Dirac band [16,46–54,58].

The above results indicate that with default *d*-orbital orientations, a Kagome lattice and hence the Kagome metal may not host a topological FB, as commonly perceived. Importantly, we found that an effective way to make the other four *d*-orbitals to conform the Kagome lattice symmetry is to rotate two of three *d*-orbitals clockwise/counterclockwise by a degree of $2\pi/3$ within a unit cell, so that they conform the three-fold rotation among the three sublattice sites A, B and C [marked in the first column of Fig. *1*(a)] plus translation in the Kagome lattice.

Starting from the default *d*-orbital orientation in Fig. 1(a), we rotate two of them clockwise/counterclockwise by $2\pi/3$, as indicated by the curved black arrows, to arrive at the configuration of the rotated *d*-orbital basis in Fig. 1(c). Remarkably, now they all produce an ideal FB, as shown in Fig. 1(d). Interestingly, one also sees that the lattice having rotated $d_{x^2-y^2}/d_{xy}$ orbitals [the second/third column in Fig. 1(d)] leads to a FB of opposite chirality sitting above/below the Dirac bands, respectively [Similarly for $d_{zx}/d_{yz}$ orbitals in the fourth/fifth column of Fig. 1(d)]. This means that due to the directional dependence of *d*-orbital hopping there are two groups of *d*-orbital Kagome lattices have effectively the lattice hopping of opposite sign ($\pm t$) [58]. Also, we emphasize that the FB resulted from the rotated *d*-orbital Kagome lattices as shown in Fig. 1(d) is symmetry protected and hence robust, independent of variations of hopping strength ($V_{dd\sigma}$, $V_{dd\pi}$ and $V_{dd\delta}$).

*Compatibility of sublattice intercalation with the d-orbital Kagome lattice.—* In Kagome metals, TM Kagome lattice is usually intercalated with another sublattice. Next, we examine whether the intercalated sublattice is compatible with the rotated *d*-orbital basis. Especially we consider the hexagonal and trigonal sublattice intercalation, as shown in Fig. *2*(a) and (b), respectively. They both preserve the $C_3$+T symmetry, which is a critical condition for the existence of FB as shown above. For simplicity, assuming one *s*-orbital at each site of hexagonal or triangular sublattice, the calculated bands are shown in Fig. *2*(c) and (d) for varying onsite energy differences ($\Delta_{sd} = \varepsilon_s - \varepsilon_d$) and interaction strength ($V_{sd\sigma}$) between *s*- and *d*-orbitals. Here we show the case of $d_{x^2-y^2}$ in Fig. *2* for illustration, and the other cases are shown in Fig. S 1 and SM [59] with qualitatively the same behavior. Red and blue bands present the intercalation-sublattice and Kagome-sublattice projection, respectively. The TB bands obtained with three sets of parameters [($\Delta_{sd} = 10t_0$, $V_{ss\sigma} = -1.2t_0$, $V_{sd\sigma} = 0$ ), ($\Delta_{sd} = 0$, $V_{ss\sigma} = -1.2t_0$, $V_{sd\sigma} = 0$), and ($\Delta_{sd} = 0$, $V_{ss\sigma} = -1.2t_0$, $V_{sd\sigma} = 1t_0$)] are shown in upper, middle, and lower panel of Fig. *2*(c) and (d), respectively.

Most interestingly, with the hexagonal intercalation [Fig. *2*(c)], an ideal FB emerges all the time, consistent with the rotated *d*-orbital Kagome bands discussed above, independent of $V_{sd\sigma}$ and $\Delta_{sd}$. It would overlap with the Dirac bands formed by *s*-orbital of hexagonal sublattice when $\Delta_{sd}$ is small. In contrast, the trigonal intercalation is less effective, making the FB less flat and also mixed with other bands [Fig. *2*(d)].

*The effect of Crystal field splitting.—* In the above analyses, we consider only one *d*-orbital per Kagome lattice site, which corresponds to the limit of a large CFS so that this *d*-band is energetically well separated from other *d*-bands. In real Kagome metals, the CFS can vary, and it turns out that in addition to *d*-orbital rotation, another general condition for the existence of ideal FB is to have a strong enough CFS ($\Delta_0$) exceeding the bandwidth ($W$), namely $\Delta_0 \gg W$ (see Fig. S2 and related discussion in the SM [59]). CFS is determined by local point-group symmetry of atoms (or molecular motifs) coordinated with the center TM atom and their bonding strength. The former dictates the split *d*-level degeneracy, while the latter affects the magnitude of energy splitting. When $\Delta_0 \gg W$, the interaction between different *d*-orbitals is suppressed, so that a specific subset of *d*-orbital bands is energetically well separated from other *d*-bands so that it can be effectively rotated to form the ideal FB.

One way to tune the CFS is by changing the intercalation potential, such as by placing a single atom versus a benzene molecule at each hexagonal sublattice site of intercalation (see Fig. S3 [59]). The condition of $\Delta_0 \gg W$ is found to be satisfied by the benzene-molecule intercalation with a strong ligand field but not by single-atom intercalation. These are indeed confirmed by real-material calculations, as demonstrated below.

*Comparison between inorganic and MOF Kagome metal.*—TM Kagome lattice are often found in two material systems, inorganic and MOF Kagome metals. However, we found there are qualitative differences in their band structure because of different intercalation sublattices of atoms and molecules. Our analyses based on orbital projection of bands indicates that layered MOF Kagome metal is more attainable to ideal FB.

Figure *3* shows the crystal structure and DFT band structure of (a) CoSn and (b) $Ni_3C_{12}S_{12}Li_6$, chosen as the representative of inorganic and MOF Kagome metals, respectively. In CoSn, the Co *d*-orbitals on the Kagome sublattice are *effectively* rotated by in-plane triangle intercalation of Sn *s*-orbital with additional two layers of Sn above and below in hexagonal symmetry. In $Ni_3C_{12}S_{12}Li_6$, the Ni *d*-orbitals on the Kagome sublattice are *ideally* rotated by hexagonal intercalation of benzene rings bridged with S atoms, while constituent layers are stacked with interlayers of Li. Figure 3(c) and (d) are calculated band structures of CoSn and $Ni_3C_{12}S_{12}Li_6$ showing the rotated *d*-orbital projection in the same energy window (-4eV to +4eV), respectively. From the orbital projection, one sees that in $Ni_3C_{12}S_{12}Li_6$, $\Delta_0 \gg W$ is satisfied, whereas in CoSn [Fig. 3(c)], it is not. In CoSn, some *d*-bands appear flat, such as $d_{x^2-y^2}$-bands (cyan dots right below the Fermi level), but not perfectly flat, and more problematic that all five *d*-bands overlap heavily with each other. In contrast, in $Ni_3C_{12}S_{12}Li_6$ [Fig. 3(d)], one sees "isolated" ideal FBs, arising from $d_{xy}$ (red dots) and $d_{zx}$ orbitals (blue dots), near the Fermi level. They are well separated from other bands. Thus, the conditions of orbital rotation and large CFS are both met in MOF Kagome metal (see also Fig. S 4 and additional discussion in the SM [59]). We note that here the bands are purposely projected onto the rotated *d*-orbitals, instead of default orientation of *d*-orbitals usually done in previous studies [25–29] [see Fig. S 5 in SM [59]], to explicitly show the effect of *d*-orbital rotation.

One may further analyze the role of CFS in CoSn versus in $Ni_3C_{12}S_{12}Li_6$, based on their respective crystal structures. In CoSn, Co atoms on the Kagome sublattice are coordinated with six neighboring Sn atoms in an elongated octahedral cage with $D_{4h}$ point group symmetry, and the two out-of-plane Sn atoms form a triangle intercalation [Fig. 3(a)]. Overall, the CF around the Co atoms is highly symmetric, leading to a very small CFS and hence a much smaller *d*-band splitting than bandwidth. In contrast, in $Ni_3C_{12}S_{12}$, the $C_6S_6$ molecules sit at the hexagonal sublattice sites, to constitute the desired hexagonal-intercalated *d*-orbital Kagome model (Fig.

2b). Also, Ni atoms on the Kagome sublattice are coordinated with four S atoms in a planar rectangular geometry, leading to a large CFS with $D_{2h}$ point group symmetry for $\Delta_0 \gg W$, in agreement with our model analysis (Fig. S2 and S3). Consequently, ideal FB emerges independent of *s-d* and *d-d* hopping strength. Moreover, the Li intercalation makes the interlayer interaction between the $Ni_3C_{12}S_{12}$ Kagome planes rather weak, as reflected by a small band dispersion along the out-of-plane direction.

*Conclusion.*—We have developed a TB model to reveal the importance of *d*-orbital rotation and strong CFS to form ideal topological FBs in Kagome metals. In intercalated Kagome metals, the hexagonal intercalation sublattice is found most compactible with the *d*-orbital rotation and a sizable intercalating molecule with strong ligand field is favorable for increasing CFS. We have examined the TM coordination and local symmetry in the reported inorganic Kagome metals (see Fig. S6, S7 and Table S3 [59]), and in all cases, the *d*-orbital is effectively rotated but the CFS is not large enough for ideal FB. We propose a family of layered MOF Kagome metals as the best candidate materials for realizing ideal FB.

This work is supported by U.S. DOE-BES (Grant No. DE-FG02-04ER46148). Support from the CHPC at the University of Utah and the NERSC at the Office of Science in the U.S. Department of Energy is gratefully acknowledged.

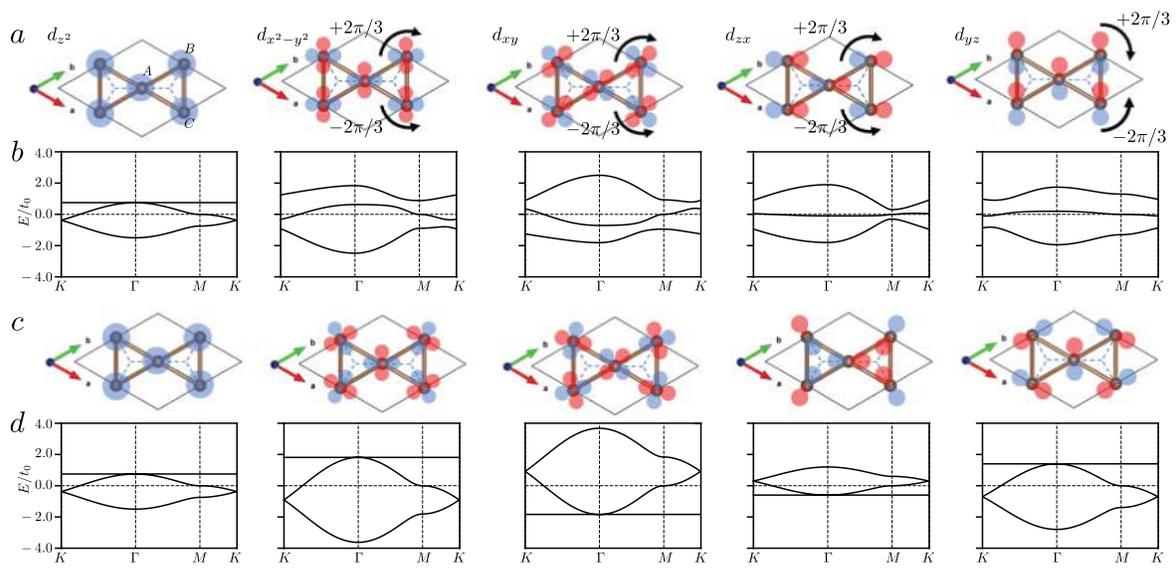

Figure 1. [Double-column] Single *d*-orbital rotation in Kagome lattice to form ideal FBs. (a) Conceptual diagram of single *d*-orbital with default orientation ($d_{z^2}$, $d_{x^2-y^2}$, $d_{xy}$, $d_{zx}$, and $d_{yz}$) and corresponding (b) TB band structures. (c) Conceptual diagram of the rotated *d*-orbital basis [indicated by black arrows in (a)] and corresponding (d) TB band structures. Red and blue (balloon) represents the positive and negative nodes of *d*-orbital wavefunctions.

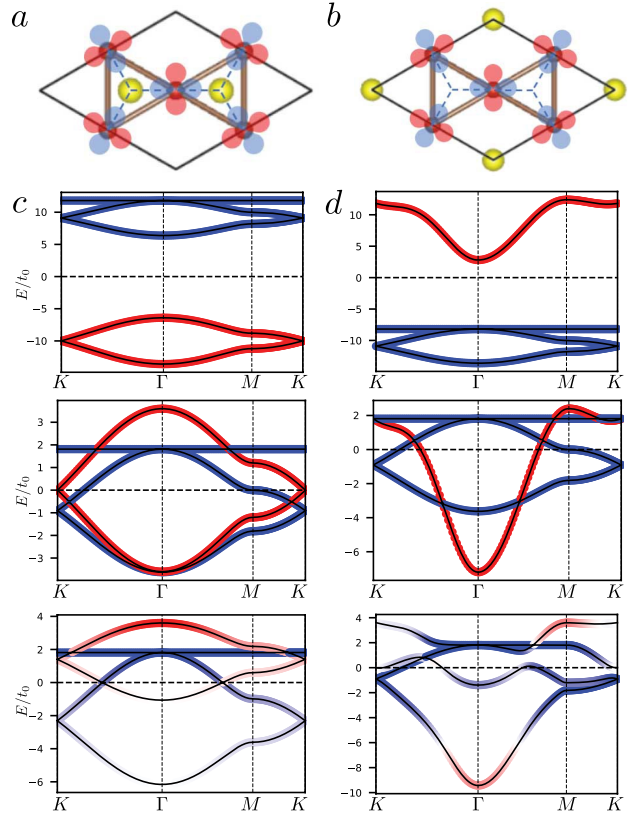

Figure 2. [Single-column] FBs in intercalated Kagome lattice. Schematic crystal structures of intercalated $d_{x^2-y^2}$-orbital Kagome lattice with *s*-orbital (a) hexagonal sublattice and (b) triangular sublattice. TB band structures of Kagome lattice intercalated with (c) hexagonal sublattice and (d) triangular sublattice. The three panels in (c) and (d) show bands with varying on-site energy difference ($\Delta_{sd}$) and *s-d* hopping integrals ($V_{sd\sigma}$); the upper, middle, and lower panels present the cases for ($\Delta_{sd} = 10t_0$, $V_{sd\sigma} = 0$), ($\Delta_{sd} = 0$, $V_{sd\sigma} = 0$), and ($\Delta_{sd} = 0$, $V_{sd\sigma} = 1t_0$), respectively. Blue and red dots represent the orbital projection onto the rotated *d*-orbital Kagome and *s*-orbital intercalated sublattice, respectively.

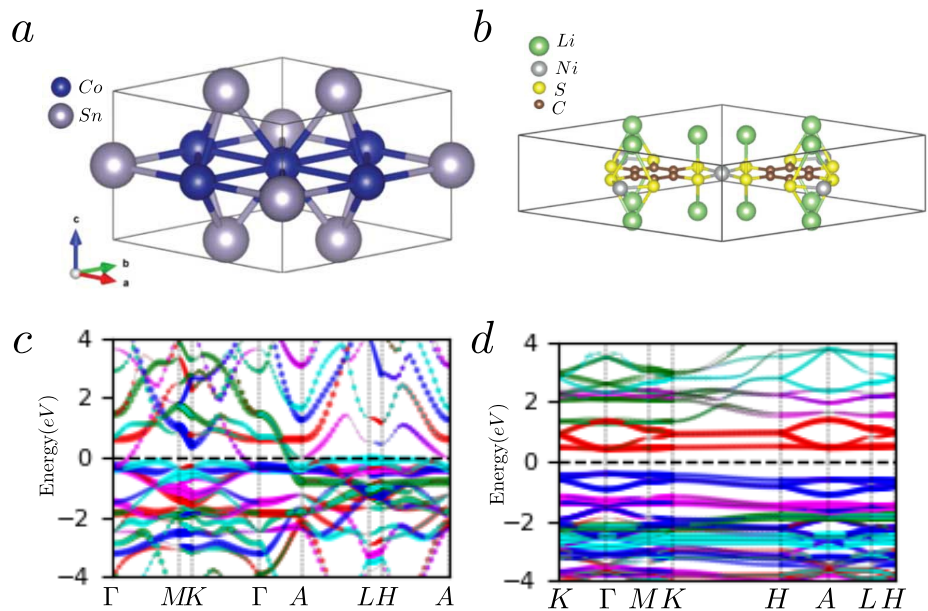

Figure 3. [Single-column] Comparison between Kagome sublattice bands in inorganic and MOF Kagome metal. (a, b) Crystal structure of CoSn and Li-intercalated $Ni_3C_{12}S_{12}$. (c) and (d) DFT band structure of CoSn and $Ni_3C_{12}S_{12}Li_6$, respectively. Color scheme in (c) and (d) represents the atomic orbital projection onto the rotated *d*-orbitals of $d_{z^2}$, $d_{x^2-y^2}$, $d_{xy}$, $d_{yz}$, and $d_{zx}$ colored in green, cyan, red, magenta, and blue, respectively.